\documentclass[12pt]{article}
\usepackage{amsmath,amsthm,amsfonts,graphicx,epsfig}
\usepackage[usenames]{color}
\numberwithin{equation}{section}

\newtheorem{thm}{Theorem}

\newcommand{\dbar}{\kern-.1em{\raise.8ex\hbox{ -}}\kern-.6em{d}}

\def\half{\mbox{$1\over2$}}
\def\quarter{\mbox{$1\over4$}}
\def\sl{SL(2,\mathbb{C})}

\def \be{\begin{equation}}
\def \ee{\end{equation}}

\author{J. E. Avron, G. Bisker and O. Kenneth
\\
Department of Physics\\ Technion, 32000 Haifa, Israel}

\begin{document}
\title{Visualizing Two Qubits}
\date{\today}%
\maketitle
\begin{abstract}
The notions of entanglement witnesses, separable and entangled
states for two qubits system can be visualized in three dimensions
using the SLOCC equivalence classes. This visualization preserves
the duality relations between the various sets and allows us to
give ``proof by inspection'' of a non-elementary result of the
Horodeckies that for two qubits, Peres separability test is
iff. We then show that the CHSH Bell inequalities can be
visualized as circles and cylinders in the same diagram.
This allows us to give a geometric proof of yet another result of
the Horodeckies, which optimizes the violation of the CHSH Bell
inequality.  Finally, we give numerical evidence that, remarkably,
allowing Alice and Bob to use three rather than two measurements
each, does not help them to distinguish any new entangled SLOCC
equivalence class beyond the CHSH class.
\end{abstract}

\section{Introduction}
The world of 2 qubits is the simplest setting where the notions of
entanglement \cite{Horodecki:review,Bruss:review,Nielsen-Chuang},
Bell inequalities \cite{Peres:book,Peres:bell} and their witnesses
\cite{Horodecki:iff}, first appear.
It would be nice if, like the Bloch sphere for one qubit \cite{Nielsen-Chuang},
they could
also be visualized geometrically. However, the world of two qubits
is represented by $4\times 4$ Hermitian matrices and being 16
dimensional it is not readily visualized. It can, however,
be visualized by introducing an appropriate equivalence relation.
This idea has been used in \cite{Leinaas} to describe the separable
and entangled states. Here we show that the entanglement witnesses
and the CHSH Bell inequalities can be incorporated in this
descriptions as well. The geometric description is faithful to the
duality between separable states and witnesses as we shall
explain. This allows for elementary and elegant proofs of
non-elementary results.

Any $4\times 4$ hermitian operator $W$ can be represented by a
$4\times 4$ real matrix $\omega$ using the Pauli matrices as the basis:
\begin{equation}\label{observable}
 W=\omega_{\mu\nu}\sigma^\mu\otimes\sigma^\nu\
\end{equation}
Greek indices run on $0,1,2,3$, Roman indices on $1,2,3$. $\sigma^0$
is the identity and $\sigma^j$ are the Pauli matrices. Summation
over a pair of repeated indices is always implied, and indices are
raised and lowered using the Minkowski metric tensor
$\eta=diag(1,-1,-1,-1)$. To reduce the number of components from 16
to 3 one relies on notions of equivalence. In particular, forgetting
about the overall normalization of operators reduces the dimension
by 1.

An effective notion of equivalence comes from allowing Alice and
Bob to operate on their respective qubits
\begin{equation}\label{eq:equivalence-states}
\rho \to \rho^M= M \,\rho\,M^\dagger,\quad M={A\otimes B},
\end{equation}
We shall focus on the case $A,B\in\sl$ where the operation is
invertible but not trace preserving. The physical
interpretation of this is that
states which are accessible by local, reversible filtering
are identified. It is known as SLOCC \cite{Bennet:SLOCC,Cirac:SLOCC} and is briefly reviewed
in section \ref{SLOCC}. Since $\dim \sl=6$ the SLOCC equivalence
reduces the dimension by 12. As a consequence, the SLOCC
equivalence classes of unnormalized 2 qubits states can be visualized in 3 dimensions.

\begin{figure}[ht]{\centering
  \includegraphics[width=8cm]{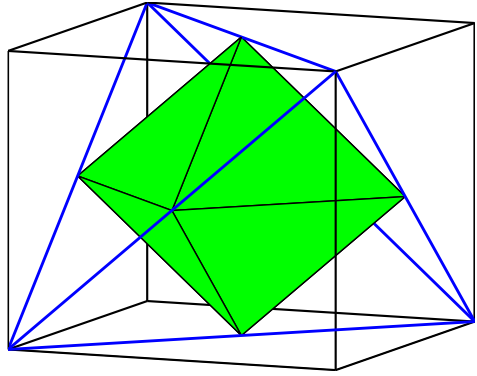}\\
  \caption{Three dimensional view of the world of two qubits:
  The cube represents the equivalence classes of potential entanglement witnesses,
  the tetrahedron represents the states, and the octahedron represents the
  separable states.
  }\label{states3D}}
\end{figure}

As we shall see, the SLOCC equivalence classes of entanglement
witnesses are represented by the cube, the states by the tetrahedron
and the separable states by the octahedron of Fig.~\ref{states3D}.
The octahedron and tetrahedron have been identified as the SLOCC
representation in \cite{Leinaas,Verstraete:Lorentz_svd}. Adding the
cube as a representation of the SLOCC equivalence classes of
entanglement witnesses, shows that the natural duality relation
between witnesses and separable states is preserved in the
visualization of the SLOCC equivalence classes: The cube is the dual
of the octahedron in the usual sense of duality of convex sets
\cite{Rockafellar}. In particular, the number of faces in one is the
number of vertices in the other. The tetrahedron is, of course, its
own dual.

Since the work of the Horodeckis,
\cite{Horodecki:information},  Fig.~\ref{states3D} has been widely
used in quantum information theory for the special cases of states
with maximally mixed subsystems \cite{Bertlmann,Werner:Entanglement-measures-symmetry}.
This is a 9 dimensional family of states
with $\omega_{0j}=\omega_{j0}=0$ in Eq.~(\ref{observable}). Since
this family has lower dimension, it can be visualized in 3
dimensions using a more restrictive notion of equivalence than
SLOCC: Alice and Bob are allowed to perform only unitary
operations on their respective qubits with $A,B\in SU(2)$ in
Eq.~(\ref{eq:equivalence-states}). This in arguably the most fundamental
notion of equivalence in quantum information theory and is
known as LOCC \cite{Nielsen-Chuang,Horodecki:review}.
It is trace preserving, which
expresses the fact that, unlike SLOCC, it is not lossy, (no state is ever
discard). Since $\dim SU(2)\times SU(2)=6$ the LOCC equivalence
classes of this  9 dimensional family of states can be represented
in 3 dimensions \cite{Horodecki:information}. It is remarkable
that both the visualization and the interpretation
of Fig.~\ref{states3D} remains the same
when one goes from the 9 dimensional family to the 16 dimensional
family of general 2 qubits states.
All that changes is the notion of equivalence.

Fig.~\ref{states3D} turns out to play a significant role also in the
theory of quantum communication.  Namely,  it characterizes the
stochastic properties of certain {\em single qubit} quantum channels
as shown in \cite{Ruskai,Ruskai:king,Ruskai:cp-trace}. This rather
different interpretation of the figure follows from a deep relation,
known as the Choi-Jamiolkwosky isomorphism \cite{Choi}, between
linear operators acting on the Hilbert space of Alice and Bob, and
linear maps on single qubit states.
Using this, one finds,
\cite{Ruskai,Ruskai:king,Ruskai:cp-trace} that (for unital and trace
preserving channels), the octahedron represents channels that destroy
entanglement, the tetrahedron represents the completely positive
maps and the cube the positive maps.

In section \ref{cto} we shall review the SLOCC interpretation of
Fig.~\ref{states3D} from a perspective that focuses on the duality
relations between the sets in the figure. The main new results
concern the visualization of entanglement witnesses, duality and
of the CHSH Bell inequalities in sections \ref{sec:vis-bell}.

\section{Local operations}\label{SLOCC}

The local mapping of a two qubit state $\rho$ given by
Eq.~(\ref{eq:equivalence-states})  preserves positivity and takes
a product state $\rho_A\otimes\rho_B$ to a product state. It
therefore maps any separable state---a convex combination of
product states---to a separable state. This makes the equivalence
$\rho\sim\rho^{M}$, a useful notion in studying the entanglement
of two qubits \cite{Leinaas,Verstraete:Lorentz_svd}. Since the operation does
not preserve the normalization of the state it is convenient to
consider states up to normalization. The operations performed by
Alice and Bob can be interpreted  as {\em probabilistically
reversible filtering} associated with the POVM
    \be
    E_1^{(M)}=\frac {M^\dagger M}{||M||^2}, \quad E_2^{(M)}=1-E_1^{(M)}
    \ee
($E_2^{(M)}$ is not a local operator.  Local POVM would require four
$E_i$'s.)
The probability of successfully filtering the state
$\rho^M/Tr(\rho^M)$ is strictly positive and is given by $Tr(\rho
E_1^{(M)})>0$. When $M$ is unitary the filtering succeeds with
probability one and no state is lost. The filtering is probabilistically reversible since the
original state $\rho$ can be recovered, with non-zero probability,
from $\rho^M$ using the filter $E_1^{(M^{-1})}$.

One can broaden the notion of equivalence under SLOCC from states to
observable and in particular, to witnesses $W$. We take the action
on witnesses to be contragradient to that of states:
\begin{equation}\label{eq:equivalence-observables}
W \to W^{M}= (M^\dagger)^{-1} \,W\,M^{-1},\quad M= A\otimes B, \quad
A,B\in \sl
\end{equation}
(In case $M$ is unitary states and observables transform the
same way). The motivation for this choice is to have $\rho W$
transforms by a similarity transformation so its trace and
therefore the associated expectation and probability, is left
invariant.

\subsection{Potential witnesses}
We shall say that $W_e$ is a {\em potential entanglement witness}
if\footnote{For a definition of witnesses that goes through the
Choi-Jamiolkowsky isomorphism, see e.g. \cite{Terhal:detecting}.}
\begin{equation}\label{ent-witness}
Tr (W_e\rho_s)\ge 0
\end{equation}
for all separable states $\rho_s$. Since the set of separable
states $\{\rho_s\}$ is a convex cone in the space of $4\times 4$
matrices, the set of potential entanglement witnesses $\{W_e\}$ is
the dual convex cone of $\{\rho_s\}$. The set of states,
$\{\rho\}$, is a convex cone as well, and the three convex cones
are evidently nested
\begin{equation}\label{nesting}
    \{\rho_s\}\subset\{\rho\}\subset\{W_e\}
\end{equation}
The cones $\{\rho\}$, $\{\rho_s\}$ and $\{W_e\}$ all lie in 16
dimensions, which is not very useful for visualization.

SLOCC takes a potential entanglement witness to a potential
entanglement witness. This follows from
\begin{equation}\label{covariance}
Tr (W_e^{M}\rho_s)=Tr (W_e\rho_s^{M^{-1}})\ge 0
\end{equation} 
which shows that $W_e^M$ is an entanglement witness if $W_e$ is.

SLOCC allows one to reduce the study of states, separable states,
and (potential) entanglement witnesses to the study of the
corresponding equivalence classes.
As will be explained in section \ref{cto},
the equivalence classes of the three cones are described by the three polyhedra
shown in Fig.~\ref{states3D}.

Similar ideas can be used to visualize Bell inequalities as we now proceed to show.


\subsection{Bell witnesses}
Every Bell inequality has a corresponding witness \cite{Terhal}.
Let $W_B$ be a witness for a specific Bell inequality. A state
$\rho_h$ satisfies this Bell inequality if
\begin{equation}\label{bell}
Tr (W_B \rho_h)\ge 0
\end{equation}
Since the separable states satisfy all types of Bell inequalities
\cite{Terhal,Horodecki:review} it is clear that $W_B$ belongs to the
family of potential entanglement witnesses. The set $\{\rho\}_B$ of
states satisfying (\ref{bell}) for a specific type of Bell
inequalities (e.g. the CHSH family) forms a convex cone. However in
general it is larger then the cone $\{\rho_s\}$ of separable states.

\subsection{CHSH witnesses}\label{sec:chsh}

 The CHSH Bell inequalities describe a situation where Alice may
choose to measure her qubits in one of two directions, $(a,a')$ and Bob may
similarly choose one of the directions, $(b,b')$.
It is represented by the witness
\cite{Terhal}
\begin{equation}\label{eq:chsh-witness}
W_B   = \half\left( 2 \pm B_{CHSH}  \right)
\end{equation}
where $B_{CHSH}$ is the CHSH operator \cite{Clauser:CHSH}:
\begin{equation}\label{bchsh} B_{CHSH}(a,a',b,b')=
 a \cdot \sigma  \otimes \left( { b +  b'} \right) \cdot
\sigma  +  a' \cdot \sigma  \otimes \left( { b -  b'} \right)
\cdot \sigma
\end{equation}
$a\cdot\sigma=a_j\sigma^j$.
The CHSH Bell inequality then
takes the form of Eq.~(\ref{bell}).

The family of all CHSH witnesses is an a-priory 8 dimensional
family associated with the 4 directions Alice and Bob choose.
(In fact,
an implicit degeneracy in
Eq.~(\ref{bchsh}) makes it is only 7 dimensional.)
LOCC takes a CHSH witness corresponding to the directions $(a,a',b,b')$
to a witness associated with rotated directions while keeping $a\cdot a'$
and $b\cdot b'$ fixed.  This reduces the dimension by 6 and allows us to
visualize the equivalence classes of CHSH witnesses.
As we explain in section \ref{sec:vis-bell},
they turn out to be the three circles shown in Fig. \ref{bell_picture}.

\subsection{Bell inequalities and SLOCC}
\color{black}
Local operations, Eq.~(\ref{eq:equivalence-states}), are guaranteed
to take separable states to separable states. This reflects the fact
that no entangled state can ever be (locally) filtered from a separable state.
This is not the case for states that satisfy Bell inequalities. In
fact, \cite{Gisin:filters} gave examples of states satisfying all
the CHSH inequalities whose filtration violate the inequality. This is
consistent
because the positivity of $Tr (W_B \rho_h)$
{\em does not} imply positivity of $Tr (W_B \rho_h^{M})$.
SLOCC does not act nicely on CHSH. This can also be seen from
the fact that the family of CHSH witnesses {\em is not} mapped on itself
by SLOCC. It is therefore {\em not possible} to represent the states
that satisfy CHSH inequalities in terms of their SLOCC equivalence
classes.

It seems clear however that if a state $\rho^M$, filtered
from $\rho$ (using only local operations), breaks a certain Bell inequality
then the properties of $\rho$
itself are inconsistent with (more general)
local hidden variables. Indeed if $\rho$ would have
been describable in terms of hidden variables that would have
implied that the results of any experiment done on it including one
involving local filtration (which is also a type of measurement)
should be explainable in terms of these hidden
variables\footnote{It turns out the combined experiment done on
$\rho$ consisting of filtration plus subsequent spin measurement can
be described using hidden variables only if the choice of Alice
whether to measure $a$ or $a'$ can influence the result of the
filtration which was completed prior to it. This breaks natural
causality assumptions. The fact that $\rho$ does not itself break
Bell's inequality may then be traced to the fact that the standard
derivation of Bell's inequalities does not involve these causality
assumptions.}.

This motivates the introduction of a notion of states that satisfy Bell
inequality in a SLOCC sense by requiring not only the state
$\rho_h$ satisfy the Bell inequality $W_B$, but also that all
states that can be {\em probabilistically filtered} from $\rho_h$
do. This is clearly a SLOCC invariant notion. Mathematically, it
is expressed by the requirement
\begin{equation}\label{strong-equivalence}
\inf_{M} Tr (W_B \rho_h^{M})\ge 0
\end{equation}
We shall denote this set $\{\rho\}_B^{SLOCC}$. It is, of course, a
smaller set, $\{\rho\}_B^{SLOCC}\subset\{\rho\}_B$, but it is not
empty.
This set is SLOCC invariant and so visualized in three dimensions.
The corresponding 3-dimensional set is the intersection of $3$ cylinders shown
in Fig.~\ref{bell3D}, as we shall explain in section \ref{sec:vis-bell}.
%
%
For any fixed witness $W_B$ the equivalence class $\{\rho\}_B^{SLOCC}$ is a convex
cone since
    \be
    \inf_{M} Tr (W_B (\rho_1+\rho_2)^{M})
    \ge \inf_{M} Tr (W_B \rho_1^{M})+ \inf_{M} Tr (W_B \rho_2^{M})
    \ee
Since the intersection of convex cones is a convex cone, it follows
that the states that satisfy {\em a family} of Bell inequalities in
the SLOCC sense also form a convex cone. In particular, this is so
for the CHSH family. The intersection of the cone with the
hyperplane $Tr\rho =1$ is then evidently a convex set.

Similarly, the dual (convex)
cone to $\{\rho\}_B^{SLOCC}$  is SLOCC invariant by
Eq.~(\ref{covariance}), i.e. if $W_B$ is a witness for a given SLOCC
family, so is $W_B^M$.
These notions of witnesses and states conform
to the notion of SLOCC. They can
therefore be represented in terms of their
equivalence classes and can be visualized in three dimensions.

\section{Lorentz Geometry of Two Qubits}\label{cto}

To describe the SLOCC equivalence classes of qubits it is convenient
to use their Lorentz description. Any single qubit observable $Q$
can be written as
\begin{equation}\label{single-qbit}
    Q=q_\mu\sigma^\mu
\end{equation}
The observable $Q$ is then represented by the real 4-vector $q$.

$Q$ is positive, and so is a state, if its trace and determinant are positive.
Since $Tr Q= 2 q_0>0$ and $\det Q=q_\mu q^\mu\ge 0$, states are
described by 4-vectors $q$ that lie in the forward
light-cone.         Consider
    \begin{equation}
    Q^M=MQM^\dagger=q_\mu \,(M\sigma^\mu M^\dagger), \quad M\in\sl
    \end{equation}
Since $q_\mu q^\mu=\det Q= \det Q^M$, it follows that the action of $M\in \sl$
on the observable $Q$ can be implemented by an (orthochronous)
Lorentz transformation of the four vector $q$. Namely,
\cite{Tung},
\begin{equation}\label{lorentz}
Q^M=q_\mu \,(M\sigma^\mu M^\dagger)= (\Lambda_M q)_\mu\sigma^\mu ,
\quad \Lambda_M\in SO_+(1,3)
\end{equation}
Similarly, any observable $W$ in the world of 2 qubits can be
represented by by a $4\times 4$ real matrix $\omega$ as in
Eq.~(\ref{observable}). This representation allows a simple
geometric characterization of potential entanglement witnesses
in terms of matrices $\omega$ that map the forward light-cone
into itself. This follows from the fact that
$W$ is an entanglement witness iff for any product state,
represented by
time-like vectors $\rho_a,\ \rho_b$:
\begin{equation}\label{forward-lc}
    Tr (W\rho_a\otimes\rho_b)= 4 \omega_{\mu\nu}\rho_a^\mu\rho_b^\nu
    =4\rho_a^\mu(\omega \rho_b)_\mu \ge 0
\end{equation}
This characterization of potential witnesses will play a role in what follows.

To describe the SLOCC equivalence classes we shall consider invariants
under the action (\ref{eq:equivalence-states}). The pair $A,B\in\sl$
associated with $M=A\otimes B$ gives rise to a pair of Lorentz
transformations $\Lambda_A$ and $\Lambda_B$ such that
\cite{Verstraete:Lorentz_svd}:
\begin{equation}\label{lorentz-2}
    \omega^M=\Lambda_A\omega\Lambda_B^T
\end{equation}
Since  $\det \Lambda_A=\det \Lambda_B=1$, $\det \omega$ is an
invariant.

\color{black}

A more interesting and powerful invariant is constructed as
follows: The Minkowski adjoint of $\omega$ is defined as:
\begin{equation}
\omega^\star=\eta \omega^T \eta
\end{equation}
where $\eta$ is the Minkowski metric tensor. $\omega^\star$
transforms contragradiently to $\omega$ under $M$. This follows
easily from the defining relations of the Lorentz transformation
$\Lambda \eta \Lambda^T=\eta$:
\begin{eqnarray}\label{onega-star-m}
(\omega^M)^\star&=&\eta (\omega^M)^T \eta \nonumber \\
&=& \eta(\Lambda_A\omega \Lambda_B^T)^T\eta \nonumber \\
&=& (\eta\Lambda_B\eta)(\eta\omega^T\eta)(\eta\Lambda_A^T\eta) \nonumber \\
&=& (\Lambda_B^T)^{-1}\omega^\star (\Lambda_A)^{-1}
\end{eqnarray}
It follows that $\omega^\star\omega$ undergoes a similarity
transformation under the action of $M$, so its spectrum is a SLOCC
invariant.

For a general observable, $\omega^\star \omega$ is not guaranteed to
be a hermitian matrix, and its spectrum therefore may not be real.
However, if $W$ is a potential entanglement witness, a
simplification occurs. In particular, the eigenvalues of
$\omega^\star \omega$ are guaranteed to be positive
\cite{Verstraete:Lorentz_svd,Leinaas}. This can be seen from the following
argument: Suppose $W$ is a {\em strict} witness of entanglement, so
that Eq.~(\ref{forward-lc}) holds with strict inequality.
$\omega^\star\omega$ then maps the forward light-cone into its
interior, since for any causal vector $\rho$
    \be
    0<(\omega \rho)_\mu(\omega \rho)^\mu=\rho_\mu (\omega^\star\omega
    \rho)^\mu
    \ee
This implies, by a fixed point argument, that the largest
eigenvalue of  $\omega^\star\omega$ is positive and has a
time-like eigenvector. The Lorentz orthogonal subspace to this
eigenvector is a space-like invariant subspace. Restricted to
this subspace, the Minkowsky adjoint coincides with the ordinary
adjoint. This makes the remaining eigenvalues positive as
well\footnote{If one replaces $<$ by $\le$ above then it might
happen that the largest eigenvalue has an eigenvector which is
light-like. This case is much more complicated, but for our aims
here can be handled by a limiting argument.}.

Up to sign the {\em Lorentz singular values} \cite{Verstraete:Lorentz_svd} are
defined as the roots of the eigenvalues of $\omega^{\star}\omega$.
We denote them by   $\omega_\alpha$.
They are the Lorentz analog of the singular
values of a matrix\footnote{$\omega_\alpha$ are not
the covariant components of a Lorenzian 4-vector.}.
As the above argument shows the largest singular
value which will be denoted $\omega_0$
corresponds generically to a timelike eigenvector (and in degenerate cases to a  null one)
while $\omega_1,\omega_2,\omega_3$ generically correspond to a spacelike eigenvector
(or possibly a null one in degenerate cases).

Defining $\omega_\alpha$ as the square roots of the (necessarily
positive) eigenvalues of $\omega^{\star}\omega$ still leaves a sign
ambiguity.
A unique determination $\omega_\alpha$ is achieved by letting
$\omega_3$ take the sign of $\det \omega$ and choosing all others
non-negative. Furthermore, one orders them according to
\begin{equation}\label{ww}
\omega_0\ge \omega_1\ge \omega_2\ge  |\omega_3|,\;\; sign(\omega_3)=sign(\det\omega)
\end{equation}
Unless $\omega_0$ happens to vanish the SLOCC equivalence class may be
characterized, up to scaling, by the three vector
    \be \label{coord}
   \vec\omega= \frac 1 {\omega_0}\big(\omega_1, \omega_2, \omega_3\big)
   \ee
 The SLOCC equivalence classes of potential entanglement witnesses would
then be represented by the pyramid
    \be\label{eq:pyramid}
    \{(x,y,\pm z)\  |\  x\ge y\ge
    z\ge 0 \}
    \ee

Remarks:

\begin{itemize}
\item If $W$ is a strict potential witness (implying $\omega_0>|\omega_i|$)
then it turns out \cite{yosi_oded} that
in analogy to the usual singular value decomposition one can find a pair of
Lorentz transformations that bring $\omega$ to its {canonical form}
\begin{equation}\label{canonical-form}
\sum\omega_\alpha \sigma^\alpha\otimes\sigma^\alpha
\end{equation}
This in turn imply that the singular values $\omega_\alpha$
completely determine $W$'s SLOCC equivalence class.
\item However if $\omega_0=max(|\omega_i|)$ (corresponding to a non-strict potential witness)
then there is no a-priori guarantee that such a pair of Lorentz transformations exist.
Witnesses $W$ associated with this boundary case split into nonequivalent classes:
those having the canonical form (\ref{canonical-form}) and others
having more complicated canonical forms \cite{Verstraete:Lorentz_svd}.
Thus in the boundary case $\omega_\alpha$ do not completely determine the
 SLOCC equivalence class.
 \end{itemize}

Note that the condition
$\omega_0\geq|\omega_1|,|\omega_2|,|\omega_3|$ is enough to
guarantee that $\omega=diag(\omega_0,\omega_1,\omega_2,\omega_3)$
takes the forward light-cone to itself and hence to guarantee that a
potential entanglement witness
$W=\sum\omega_\alpha\sigma^\alpha\otimes\sigma^\alpha$ having
$\omega_\alpha$ as singular values exists.

\color{black}
Operators $W=\sum\omega_\alpha\sigma^\alpha\otimes\sigma^\alpha$
which differ by permutation of $\omega_1,\omega_2,\omega_3$ or by
flipping a sign of a pair of $\omega_i$'s are  SLOCC
equivalent.
There are 24 such operations corresponding to the tetrahedral
group.
Strictly, therefore, the SLOCC equivalence
classes of $W$
are represented by the pyramid of  (\ref{ww}). However for the purpose of
drawing pictures it is more aesthetic to
symmetrize and give up (\ref{ww}).
Now each (generic) equivalence class is
represented by 24 points in the $\omega_i$'s space(\ref{coord}). In
particular, the potential entanglement witnesses are then
represented by the unit cube.

\subsection{SLOCC and duality}

The following fact \cite{yosi_oded} allows one to translate the duality relation
between potential witnesses and separable state from 16 dimensions to 3:
\begin{thm}\label{oded_lemma}
Let $W$ and $W'$ be two  potential entanglement witnesses. Then:
    \begin{equation}\label{duality}
    \inf_{M,N}\ Tr (W^N W'^M)= 4  \left(\omega_0 \omega'_0-\omega_1 \omega'_1-\omega_2 \omega'_2
    +\omega_3 \omega'_3\right)
    \end{equation}
  where $\omega_\alpha$ and $\omega_\alpha'$ are the Lorentz singular
  values of $W$ and $W'$ respectively, ordered according to Eq.~(\ref{ww}).
\end{thm}


From a geometric point of view it may be more aesthetic to
use an equivalent formulation of the theorem which allows using any of the 24
possible representatives $\omega_\alpha$ (not necessarily satisfying (\ref{ww})).
This is easily achieved by replacing the r.h.s. of Eq.~(\ref{duality}) by

$$4\min \left\{
  \omega_0 \omega'_0+\omega_1 \omega'_1+\omega_2 \omega'_2
    +\omega_3 \omega'_3  \right\}$$
where the minimum is taken over the 24 possible representatives $\omega'_\alpha$.

\begin{figure}[ht]{\centering
    \includegraphics[width=8cm]{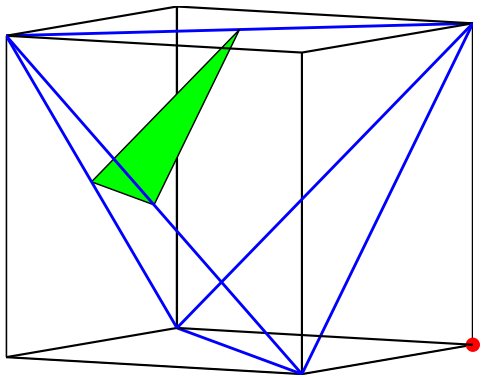}\\
    \caption{The red dot  in the lower right corner is an 
    entanglement witness associated with $\vec\omega$.  The green
    triangle lies on the plane $\vec\rho\cdot \vec\omega=-1$.
    One may think of points near the corner of the tetrahedron
    that lie beyond the green triangle as representing the states that are
    incriminated as entangled
     by $\vec\omega$. The chosen witness is  optimal in the
    sense that no other entanglement witness detects a larger
    set of  entangled states.}\label{states3D_singlet_witness}}\label{fff}
    \end{figure}

In particular, given a potential witness $W$ and a separable state
$\rho_s$, we have for any representatives as in Eq.~(\ref{coord}),
$\vec{\omega}$ and $\vec{\rho}$, the inequality
    \be\label{eq:cube-octa-duality}
    0\le 
    4(1+\vec\omega \cdot \vec \rho)
    \ee
Fig.~\ref{fff} demonstrates this inequality for a
particular choice of $W$.

Since the positivity of the right hand side is the standard duality
relation between convex sets in 3 dimensions \cite{Rockafellar} we
see that the theorem translates the duality,
Eq.~(\ref{ent-witness}), between the 16 dimensional cones, to the
duality between convex sets in 3 dimensions \cite{Rockafellar}.

Letting the tetrahedral group act on the pyramid of
Eq.~(\ref{eq:pyramid}) gives the unit cube.  Since the cube is the
dual (also known as Polar \cite{Rockafellar}) of the octahedron
one learns from Eq.~(\ref{eq:cube-octa-duality}) that the SLOCC
equivalence classes of the separable states are represented by the
octahedron (up to the tetrahedral symmetries).

The 16 dimensional set of states is self-dual. By
Eq.~(\ref{eq:cube-octa-duality}) the corresponding SLOCC equivalence classes
must be represented by a self-dual convex set in three dimensions,
which turns out to be the tetrahedron. To see this note that an
operator $\rho$ in the canonical form, Eq.~(\ref{canonical-form}), is
a sum of mutually commuting operators with one relation, ($\prod
\sigma^\mu\otimes\sigma^\mu=-1$). It follows that its eigenvalues
are
    \be \{\rho_0+\epsilon_1
    \rho_1 + \epsilon_2 \rho_2 - \epsilon_1\epsilon_2
    \rho_3\}\ee
where $\epsilon_1,\epsilon_2\in\{+1,-1\}$. Requiring $\rho$ to be
positive restricts to the intersection of four half-spaces which
evidently yield the tetrahedron. This completes the derivation of
Fig.~\ref{states3D}.

One nice consequence of the geometric construction  is a
``proof by inspection'' that for 2 qubits the Peres separability test is iff \cite{Leinaas}.
It is easy to see \cite{Peres:test} that if $\rho$ is a
separable state, then its partial transpose is positive.
The converse is not true in general, but is true for 2 qubits.
However the proof \cite{Horodecki:iff} rests on non-elementary
facts from operator algebras\footnote{See \cite{Terhal:detecting} for the history of this problem.}.

The proof by inspection goes as follows \cite{Leinaas}:
Denote by $\rho^P$ the partial transposition of $\rho$.
Since $\sigma_2$ is
antisymmetric, while the remaining $\sigma_\mu$ are symmetric, one has
     \be
     (\rho^P)_{\mu\nu}=\left\{%
    \begin{array}{rl}
    -\rho_{\mu 2}, & \hbox{for $\nu=2$} \\
    \rho_{\mu\nu}, & \hbox{otherwise} \\
    \end{array}%
    \right.
    \ee On the SLOCC equivalence classes of states
$\rho=\varrho_\alpha\sigma^\alpha\otimes\sigma^\alpha$, the partial
transposition then acts as a reflection in the 2 axis: Replacing
$\varrho_2$ with $-\varrho_2$, Since the octahedron of separable
states is the intersection of the tetrahedron with its reflection
through the $\varrho_2=0$ plane the result follows.

\section{Visualizing the CHSH inequalities}\label{sec:vis-bell}

\color{black}

The CHSH witnesses and inequalities were described in
section~\ref{sec:chsh}. For the sake of simplicity in notation we
shall now stick with the plus sign in the witness of Eq.~(\ref{eq:chsh-witness})\footnote{The
minus sign corresponds to flipping $\hat{b},\hat{b}'$.}.
\begin{figure}[ht]{\centering
  \includegraphics[width=10cm]{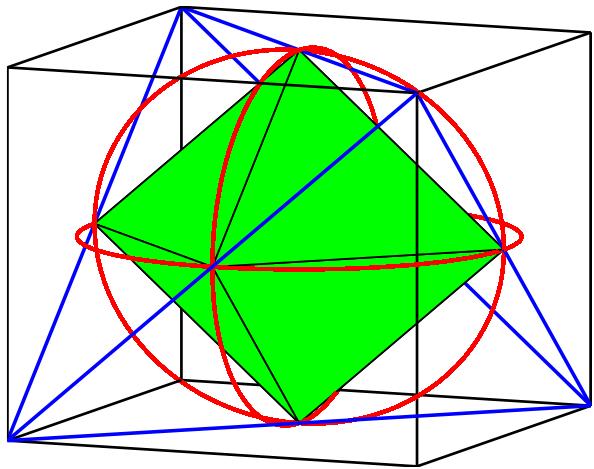}\\
  \caption{The circles represent the CHSH Witnesses}\label{bell_picture}}
\end{figure}

\color{black}

If a state violates a CHSH inequality
then it is necessarily entangled, but the opposite claim is false;
There are entangled states that do not violate any CHSH inequality.
Our aim is to visualize these.

\color{black}

The CHSH witnesses have the property that
$(\omega_B)_{0j}=(\omega_B)_{j0}=0$.
This family is invariant under the action of LOCC.
The associated equivalence classes then live in three dimensions
and are in 1-1 correspondence with the 3 singular values of
the $3\times 3$ matrix $\tilde \omega_B$
    \be\label{3by3mtarix} (\tilde \omega_B)_{ij}= (\omega_B)_{ij}\ee The
singular values are the roots of the three eigenvalues of
$\tilde\omega_B^\dagger\tilde \omega_B$.

To find the explicit dependence of the singular values on the LOCC
invariants $\cos\alpha=\hat{a}\cdot\hat{a}'$ and
$\cos\beta=\hat{b}\cdot\hat{b}'$---the angles between the two
directions Alice and Bob choose---we introduce the pair of
$3\times 2$ matrices: %
\be A=\big(\hat a,\hat a'\big)
    \quad \quad
     \quad
B=\big(\hat b,\hat b'\big)
\ee%
One checks that
    \be 2\tilde\omega_B=
a\otimes(b+b')^T+a'\otimes(b-b')^T=
A\left( \begin{array}{rr}
1 &  1 \\  1 & -1 \\
 \end{array}\right) B^T\ee
Since $\tilde\omega_B^\dagger \tilde\omega_B$ is manifestly a
$3\times 3$ matrix with rank $2$, one of its eigenvalues is zero.
Its remaining nonzero eigenvalues equal to those of the $2\times 2$
matrix
    \be\label{two-by-two}
    \quarter(B^T B) \left( \begin{array}{rr}
1 &  1 \\  1 & -1 \\  \end{array}\right)
 (A^T A)\left( \begin{array}{rr} 1 &  1 \\  1 & -1 \\
 \end{array}\right) \ee %
Evidently
    \be A^T  A=\left(
               \begin{array}{cc}
                 1 & \cos \alpha \\
                 \cos \alpha & 1 \\
               \end{array}
             \right),
             \quad
B^T  B=\left(
               \begin{array}{cc}
                 1 & \cos\beta \\
                 \cos\beta & 1 \\
               \end{array}
             \right) \quad \quad
    \ee
The matrix in Eq.~(\ref{two-by-two}) now takes the form
    \be\label{trace_one}\left(
    \begin{array}{cc}
    \cos^2(\frac{\alpha}{2}) & \cos\beta\sin^2(\frac{\alpha}{2}) \\
    \cos\beta\cos^2(\frac{\alpha}{2}) & \sin^2(\frac{\alpha}{2}) \\
    \end{array}
    \right)
\ee
It has a unit trace, so the singular values of $\tilde\omega_B$ lie
on the unit circle $\omega_1^2+\omega_2^2=1,\;\omega_3=0$. Solving
for the eigenvalues we find:
$$2\omega_{1,2}^2=1\pm\sqrt{1-\sin^2\alpha\sin^2\beta},\;\omega_3=0$$
As $\alpha$ and $\beta$ vary from $0$ to $2\pi$ this gives one
eighth of the unit circle where $1\geq\omega_1\geq\omega_2\geq0$. If
we adjoint to it the twenty four representatives of the same
equivalence class, we get the three mutually intersecting unit
circles, shown in Fig.~\ref{bell_picture}, representing the
LOCC equivalence classes of CHSH witnesses.

\begin{figure}[ht]{\centering
  \includegraphics[width=8cm]{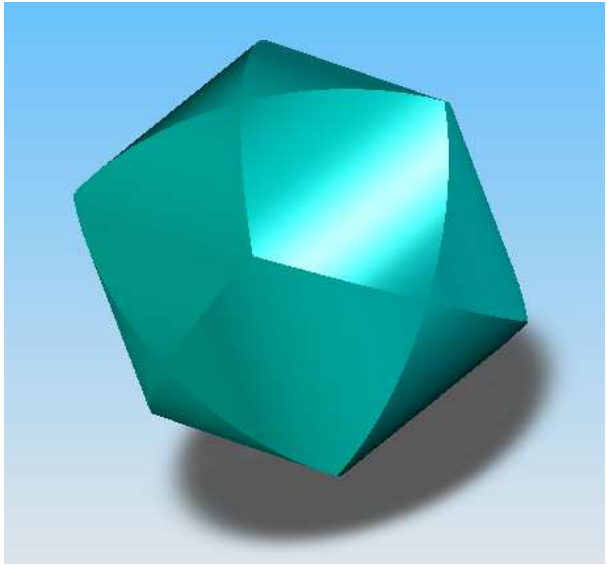}\\
  \caption{ The set of states that satisfy all CHSH
inequalities in the SLOCC sense is the intersection of three
cylinders.}\label{bell3D}}
\end{figure}

The dual set to the three unit circles
(more precisely to their convex hull)
then represents the LOCC
equivalence classes of the states that satisfy all the CHSH
inequalities\footnote{\label{f4}A-priori, only states with completely mixed
subsystems are accommodated in a 3-D LOCC diagram. However, it is
easy to see that $\rho_{0j}$ and $\rho_{j0}$ do not affect $Tr(\rho W_B)$.}.
To describe this geometrically,
note that the dual set of the unit circle in the $x-y$ plane, for
example, is the cylinder along the $z$ axis, with a unit radius.
The LOCC equivalence classes of states that satisfy all the CHSH
inequalities, is the intersection of three cylinders along the
$x$, $y$ and $z$ axes, with a unit radius. This set (see Figure
\ref{bell3D}) is bigger than the set of separable states
represented by the octahedron.
\color{black}

\color{black}
\subsection{Optimizing the CHSH Inequality}
\color{black}
In \cite{Horodecki:Violating_Bell} R. Horodecki, P. Horodecki and M.
Horodecki solved the problem of finding the optimal CHSH
witness\footnote{See also \cite{Werner:bell_entangle}.} for a given
(normalized) state $\rho$.  They show that
    \be
    \quarter Tr(\rho W_B)\ge  1-\sqrt {\rho_1^2 + \rho_2^2}
    \ee
and the inequality is saturated for an appropriate choice of
angles $\alpha,\beta$. Here $\rho_{1,2}$ are the two largest singular
values of  the $3\times 3$ matrix $\tilde \rho$ constructed from
the spatial components of the matrix elements $\rho_{\mu\nu}$ as
in Eq.~(\ref{3by3mtarix}). In particular, the state $\rho$
violates a CHSH inequality if and only if ${\rho_1^2 +
\rho_2^2}>1$.

This result can be derived, essentially by inspection, from the
geometric description of the previous section. Recall, (see
footnote \ref{f4}), that the state $\rho$ may be assumed, without loss,
to be one where the subsystems are completely mixed. The LOCC
equivalence class of $\rho$ is represented by the three singular
values of $\tilde \rho$, which we denote by $\vec\rho$. A Bell
witness, is represented by the three singular values of
$\tilde\omega_B$, which we denote by $\vec\omega_B$.
The vector $\vec\omega_B$ takes values on the three circles
in the figure. For a normalized state $\rho$
    \be
    \quarter Tr(\rho W_B)= 1+\vec\rho \cdot \vec\omega_B
    \ee
It is clear that the optimal choice of a witness (a minimizer) is
to choose the witness  $\tilde \omega_B$ so that the vector
$\vec\omega_B$ is as anti-parallel to $\vec\rho$ as possible.
(Recall that $\vec\omega_B$ is constrained to lie in one of the
principle planes.) The minimizer is then smallest entry among
    \be \{ 1-|\vec\rho\times\hat x|\, ,  1-|\vec\rho\times\hat y|\,,1-|\vec\rho\times\hat z|
    \}\ee
This reproduces the result of the Horodeckies.

\subsection{SLOCC interpretation}
\color{black}
Fig.~\ref{bell3D} also admits the following SLOCC interpretation:
The states represented by points  lying in the intersection
of the three cylinders have the property that they, and all that
can be filtered from them, satisfy all the CHSH inequalities. This
is an immediate consequence of theorem \ref{oded_lemma} which
guarantees that $Tr(\rho^{A\otimes B} W_B)$ attains its minimum
value when $\rho$ takes its canonical form.

The SLOCC equivalence classes of states that lie outside the
intersection of the cylinders have the property that they can
always be filtered to yield states that violate some CHSH
inequality.

\section{More can be less}

The CHSH inequality constrains Alice and Bob to two dichotomic
tests each. A general theory of Bell inequalities allows Alice and
Bob $n_A$ and $n_B$ tests, having $m_A$ and $m_B$ possible
results. (A geometric framework for deriving such
generalized Bell inequalities is described in \cite{Peres:bell}.)
Let $I(n_An_Bm_Am_B)$ denote a corresponding Bell inequality. The
CHSH inequality is then $I(2222)$.  Von Neumann tests on qubits
restricts the outcomes of each test to two, $m_A=m_B=2$,
however, the number of tests, $n_A,n_B$ can be arbitrary.

One naively expects that by increasing the number of tests one
might be able to incriminate
some of the entangled states that
pass the CHSH test.  Indeed, D. Collins and N. Gisin
\cite{Gisin_and_Collins} present an example of a state $\rho$ that
violates an $I(3322)$ inequality but does not violate any CHSH
inequality.

Here we shall present numerical evidence which shows that when
Bell inequalities are interpreted in the SLOCC sense then
$I(3322)$ (with three dichotomic tests), is strictly weaker than
$I(2222)$. It follows that the state found by Collins and
Gisin can be filtered to a state that
violates CHSH. 

The $I(3322)$ inequality derived by D. Collins and N. Gisin
\cite{Gisin_and_Collins} takes the form
\begin{equation}
\begin{array}{l}
  {- 2P(B_1) - P(B_2) -P(A_1)+P(A_1 B_1) +  P(A_1 B_2) +
  P(A_1 B_3) +} \\
  {P(A_2 B_1)+ P(A_2 B_2)  -P(A_2 B_3 ) +
  P(A_3 B_1) - P(A_3 B_2 ) \le   0}
 \end{array}\nonumber
\end{equation}
%
   where $P(A_iB_j)$
is the probability that when Alice chooses the
$i$th measurement and Bob chooses the $j$th measurement, they both
get the outcome $0$.

Allowing Alice and Bob three experiments to choose from gives them more freedom.
In particular, they are always free to
disconnect the third experiment. This means that the CHSH, or $I(2222)$, must be a special case of
$I(3322)$. Indeed, the CHSH $I(2222)$ inequality,
\begin{equation}
\begin{array}{l}
P( {A_1 B_1 }) + P( A_1 B_2 )
  + P(A_2 B_1 ) - P(A_2 B_2)
  - P(B_1) - P(A_1)\le  0
 \end{array}\nonumber
\end{equation}
is obtained from $I(3322)$ by Alice  disconnecting her third experiment,
$P(A_3)=0$, and Bob disconnecting his first experiment, $P(B_1)=0$.
Renaming $B_2$ and $B_3$ as $B_1$ and $B_2$
respectively gives CHSH.

Dichotomic, von Neumann, tests  of Alice and Bob are
described by projection operators, namely setting in the above
    \be
    P(A_iB_j)\rightarrow\frac{1+ a_i\cdot\sigma}2\otimes\frac{1+ b_j\cdot\sigma}2
    \ee
where $ a_i$ and $b_j$ are interpreted as directions
in a measurement of the spin projection.
Note that the case of a disconnected experiment
{\em is not} of this form: It is not a dichotomic von Neumann measurement.

The witness corresponding to
$I({3322})$  with three dichotomic von Neumann measurements for both Alice and Bob is then
\begin{equation}
\begin{array}{l}
 W_{3322}  = 4I \otimes I +
 I \otimes \left( { b_1  +  b_2 } \right) \cdot  \sigma  - \left( { a_1  +  a_2 } \right) \cdot \sigma  \otimes I \\
 \quad \quad \quad - \left( { a_1  +  a_2 } \right) \cdot  \sigma  \otimes \left( { b_1  +  b_2 } \right) \cdot  \sigma \
 -
 \left( { a_1  -  a_2 } \right) \cdot  \sigma  \otimes  b_3  \cdot  \sigma \\
 \quad \quad \quad -  a_3  \cdot  \sigma  \otimes \left( { b_1  -  b_2 } \right) \cdot  \sigma  \\
 \end{array}
\end{equation}
$Tr\left( {\rho W_{3322}}\right)<0$, implies that the state $\rho$
violates a Bell inequality.

$W_{3322}$ represents an (a-priori) 12 dimensional family of witnesses. It has
six LOCC invariant parameters, the angles
$\cos(\alpha_{ij})={a}_i\cdot {a}_j$ and
$\cos(\beta_{ij})={b}_i\cdot{b}_j$, $i\neq j
\in\{1,2,3\}$.  We can visualize this family in 3 dimensions by
representing $W_{3322}=(\omega_{3322})_{\mu\nu}\sigma^\mu\otimes\sigma^\nu$
by its canonical form under $\sl$.

\begin{figure}[ht]{\centering
  \includegraphics[width=10cm]{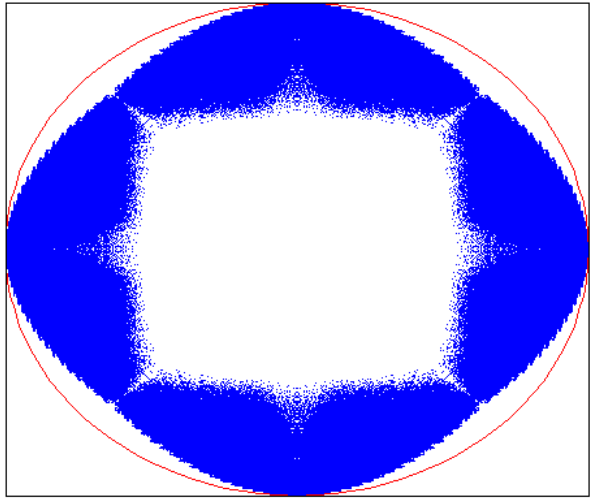}\\
  \caption{CHSH and $I_{3322}$ Witnesses in the $X-Y$ plane: the circle represent
  CHSH witnesses while all the dots inside represent $I_{3322}$ witnesses in this plane. Since the
  dots lie inside the circle they represent weaker witnesses.}\label{gisin_z}}
\end{figure}

Let us introduce the following direction matrices%
\be A=\left(
      \begin{array}{cccc}
        1 & 0 & 0 & 0 \\
        0 & {a}_{1x} & {a_2}_x & {a_3}_x \\
        0 & {a}_{1y} & {a_2}_y & {a_3}_y \\
        0 & {a}_{1z} & {a_2}_z & {a_3}_z \\
      \end{array}
    \right) \quad
    B=\left(
      \begin{array}{cccc}
        1 & 0 & 0 & 0 \\
        0 & {b_1}_x & {b_2}_x & {b_3}_x \\
        0 & {b_1}_y & {b_2}_y & {b_3}_y \\
        0 & {b_1}_z & {b_2}_z & {b_3}_z \\
      \end{array}
      \right)
\ee
 and
\be W_0=\left(
         \begin{array}{ccrr}
           4 & -1 & -1 & 0 \\
           1 & -1 & -1 & -1 \\
           1 & -1 & -1 & 1 \\
           0 & -1 & 1 & 0 \\
         \end{array}
       \right)
\ee%
so $\omega_{3322}=AW_0B^T$.  The Lorentz singular values are the
roots of the eigenvalues of  $(B^T\eta B) W_0^T (A^T \eta A)W_0$ .
One finds%
    \be A^T \eta A=\left(
               \begin{array}{cccc}
                 1 & 0 & 0 &0\\
                 0 & -1 & -\cos(\alpha_{12}) & -\cos(\alpha_{13})\\
                 0 & -\cos(\alpha_{12}) & -1 & -\cos(\alpha_{23})\\
                 0 & -\cos(\alpha_{13}) & -\cos(\alpha_{23}) & -1\\
               \end{array}
             \right)
    \ee
    \be B^T \eta B=\left(
               \begin{array}{cccc}
                1 & 0 & 0 &0\\
                 0 & -1 & -\cos(\beta_{12}) & -\cos(\beta_{13})\\
                 0 & -\cos(\beta_{12}) & -1 & -\cos(\beta_{23})\\
                 0 & -\cos(\beta_{13}) & -\cos(\beta_{23}) & -1\\
               \end{array}
             \right)
    \ee
The Lorentz singular values were calculated numerically.
The intersection of the resulting set with the $X-Y$ plane, shown in Fig. \ref{gisin_z}
is clearly contained in the CHSH unit circle.
Similarly all points outside this plane were found to lie inside the
convex hull of the three CHSH circles, implying they represent weaker witnesses.
This means that under SLOCC the CHSH inequality is stronger than $I_{3322}$ (when Alice and Bob
are constrained to measure three spin directions each).

Thus, if the two parties are allowed to filter then by
letting them choose from $3$ possible experiments, we
gain less information than by restricting them to choose from
$2$ possible experiments.

\section{Concluding remarks}
Many of the important concepts in quantum information, such as
entangled and separable states, entanglement witnesses, the CHSH
Bell inequalities etc. can be visualized in three dimensions by
introducing an appropriate equivalence relation.
The visualization allows us to give a ``proof by inspection'' of the
non-elementary fact \cite{Horodecki:iff} that the Peres separability
test for 2 qubits is iff. It also allows us to ``solve by inspection''
the problem of optimizing the CHSH Bell inequality, which was solved
by analytical methods in \cite{Horodecki:Violating_Bell}.

We have introduced the notion of states that satisfy Bell
inequalities in the SLOCC sense. We gave numerical evidence which
shows that allowing Alice and Bob an additional {\em dichotomic von Neumann} test does
not enable them to shrink the set shown in
Fig.~\ref{bell3D}, obtained by filtering and CHSH. It is an
interesting open question whether four or more dichotomic tests, or more general POVM tests, can
further shrink the set shown in Fig.~\ref{bell3D}.
\bigskip

{\bf Acknowledgment:} This work is partially supported by the ISF.
We thank Michael Burman for his help with drawing figure
\ref{bell3D}, Netanel Lindner for discussions and Mary Beth Ruskai
for a helpful correspondence and several useful suggestions.


\begin{thebibliography}{10}

\bibitem{yosi_oded}
J.~E. Avron and O.~Kenneth.
\newblock in preparation.

\bibitem{Bennet:SLOCC}
Charles~H. Bennett, Sandu Popescu, Daniel Rohrlich, John~A.
Smolin, and
  Ashish~V. Thapliyal.
\newblock Exact and asymptotic measures of multipartite pure-state
  entanglement.
\newblock {\em Phys. Rev. A}, 63(1):012307, Dec 2000.

\bibitem{Bertlmann}
R.~A. Bertlmann, H.~Narnhofer, and W.~Thirring.
\newblock Geometric picture of entanglement and bell inequalities.
\newblock {\em Phys. Rev. A}, 66(3):032319, 2002.

\bibitem{Bruss:review}
D~Bru\ss.
\newblock Characterizing entanglement.
\newblock {\em J. Math. Phys.}, 43:4237, 2002, arXive:qunat-ph/0110078.

\bibitem{Choi}
M.~D. Choi.
\newblock Completely positive linear maps on complex matrices.
\newblock {\em Lin. Alg. Appl.}, 10:285--290, 1975.

\bibitem{Clauser:CHSH}
John~F. Clauser, Michael~A. Horne, Abner Shimony, and Richard~A.
Holt.
\newblock Proposed experiment to test local hidden-variable theories.
\newblock {\em Phys. Rev. Lett.}, 23(15):880--884, Oct 1969.

\bibitem{Gisin_and_Collins}
D.~Collins and N.~Gisin.
\newblock A relevant two qubit bell inequality inequivalent to the chsh
  inequality.
\newblock {\em Journal of Physics A: Mathematical and General},
  37:1775--1787(13), 2004.

\bibitem{Cirac:SLOCC}
W.~D\"ur, G.~Vidal, and J.~I. Cirac.
\newblock Three qubits can be entangled in two inequivalent ways.
\newblock {\em Phys. Rev. A}, 62(6):062314, Nov 2000.

\bibitem{Gisin:filters}
N.~Gisin.
\newblock Hidden quantum nonlocality revealed by local filters.
\newblock {\em Physics Letters A}, 210:151--156(6), 1996.

\bibitem{Horodecki:iff}
M.~E. Horodecki, P.~Horodecki, and R.~Horodecki.
\newblock Separability of mixed states: necessary and sufficient conditions.
\newblock {\em Physics Letters A}, 223(1):1--8, November 1996.

\bibitem{Horodecki:Violating_Bell}
R.~Horodecki, P.~Horodecki, and M.~Horodecki.
\newblock Violating bell inequality by mixed spin-1/2 states: necessary and
  sufficient condition.
\newblock {\em Physics Letters A}, 200:340--344(5), 1995.

\bibitem{Horodecki:review}
R.~Horodecki, P.~Horodecki, M.~E. Horodecki, and K.~Horodecki.
\newblock Quantum entanglement.
\newblock 2007, quant-ph/0702225.

\bibitem{Horodecki:information}
Ryszard Horodecki and Michal Horodecki.
\newblock Information-theoretic aspects of inseparability of mixed states.
\newblock {\em Phys. Rev. A}, 54(3):1838--1843, September 1996.

\bibitem{Ruskai:king}
Chris King and Mary~Beth Ruskai.
\newblock Minimal enytroy of states.
\newblock {\em IEEE Trans. Info. Theory}, 47:192---209, 1999, quant-ph/9911079.

\bibitem{Leinaas}
Jon~Magne Leinaas, Jan Myrheim, and Eirik Ovrum.
\newblock Geometrical aspects of entanglement.
\newblock {\em Physical Review A (Atomic, Molecular, and Optical Physics)},
  74(1):012313, 2006.

\bibitem{Nielsen-Chuang}
Michael~A. Nielsen and Isaac~L. Chuang.
\newblock {\em Quantum Computation and Quanum Information}.
\newblock Cambridge U.P., 2000.

\bibitem{Peres:book}
Asher Peres.
\newblock {\em Quantum Theory: Concepts and Methods}.
\newblock Springer, 1995.

\bibitem{Peres:test}
Asher Peres.
\newblock Separability criterion for density matrices.
\newblock {\em Phys. Rev. Lett.}, 77(8):1413--1415, Aug 1996.

\bibitem{Peres:bell}
Asher Peres.
\newblock All the bell inequalities.
\newblock {\em Foundations of Physics}, 29(4):589--614, April 1999.

\bibitem{Rockafellar}
R.~Tyrrell Rockafellar.
\newblock {\em Convex Analysis}.
\newblock Princeton University Press, 1970.

\bibitem{Ruskai}
Mary~Beth Ruskai.
\newblock Qubit entanglement breaking channels.
\newblock {\em Rev. Math. Phys}, 15:643--662, 2003, quant-ph/0302032v3.

\bibitem{Ruskai:cp-trace}
Mary~Beth Ruskai, Stanislaw Szarek, and Elisabeth Werner.
\newblock An analysis of completely-positive trace-preserving maps on 2x2
  matrices.
\newblock 2001, quant-ph/0101003v2.

\bibitem{Terhal}
B.~M. Terhal.
\newblock Bell inequalities and the separability criterion.
\newblock {\em Physics Letters A}, 271(5):319--326, July 2000.

\bibitem{Terhal:detecting}
Barbara~M. Terhal.
\newblock Detecting quantum entanglement.
\newblock {\em Journal of Theoretical Computer Science}, 287(1):313--335, 2002,
  quant-ph/0101032.

\bibitem{Tung}
Wu-Ki Tung.
\newblock {\em Group theory in physics}.
\newblock World Scientific, 1985.

\bibitem{Verstraete:Lorentz_svd}
Frank Verstraete, Jeroen Dehaene, and Bart De~Moor.
\newblock Lorentz singular-value decomposition and its applications to pure
  states of three qubits.
\newblock {\em Phys. Rev. A}, 65(3):032308, Februar 2002.

\bibitem{Werner:Entanglement-measures-symmetry}
K.~G.~H. Vollbrecht and R.~F. Werner.
\newblock Entanglement measures under symmetry.
\newblock {\em Phys. Rev. A}, 64(6):062307, 2001.

\bibitem{Werner:bell_entangle}
Reinhard~F. Werner and Michael~M. Wolf.
\newblock Bell inequalities and entanglement.
\newblock arXiv:quant-ph/0107093, {2001}.

\end{thebibliography}


\end{document}